\documentclass[useAMS,usenatbib]{mn2e}

\usepackage{graphicx}
\usepackage{amsmath}
\usepackage{amssymb}
\usepackage{color}
\graphicspath{{figures/}}

\renewcommand{\mp}{m_\mathrm{p}}
\renewcommand{\sun}{\odot}

\renewcommand*{\vec}[1]{\textbf{#1}}

\title[3D Jet -- Disc Interaction]{
Asymmetries in Extragalactic Double Radio Sources:
Clues from 3D Simulations of Jet -- Disc Interaction}
\author[V. Gaibler, S. Khochfar, M. Krause]{
    V. Gaibler$^{1}$\thanks{E-mail: vgaibler@mpe.mpg.de}, 
    S. Khochfar$^{1}$ and
    M. Krause$^{1,2}$\\
    $^{1}$Max-Planck-Institut f\"ur extraterrestrische Physik, Gie\ss{}enbachstra\ss{}e,
    85748 Garching, Germany\\
$^{2}$Universit\"ats-Sternwarte, Ludwig-Maximilians-Universit\"at (USM),
    Scheinerstr.~1, 81679 M\"unchen, Germany}
\begin{document}

\date{Accepted ???. Received ???}

\pagerange{\pageref{firstpage}--\pageref{lastpage}} \pubyear{???}

\maketitle

\label{firstpage}

\begin{abstract}
Observational and theoretical studies of extragalactic radio sources have
suggested that an inhomogeneous environment may be responsible for observed arm
length asymmetries of jets and the properties of extended emission line regions
in high redshift radio galaxies. We perform 3D hydrodynamic simulations
of the interaction of a powerful extragalactic bipolar jet with a disc-shaped
clumpy interstellar medium of log-normal density distribution and analyze the
asymmetry. Furthermore, we compute the relation between jet asymmetry and
the ISM properties by means of Monte Carlo simulations based on a 1D
propagation model for the jet through the dense medium. We find that the
properties of the ISM can be related to a probability distribution of jet arm
length asymmetries: Disc density and height are found to have the largest
effect on the asymmetry for realistic parameter ranges, while the Fourier
energy spectrum of the ISM and turbulent Mach number only have a smaller effect.
The hydrodynamic simulations show that asymmetries generally may be even
larger than expected from the 1D model due to the complex interaction of the
jet and its bow shock with gaseous clumps, which goes much beyond simple energy
disposal. From our results, observed asymmetries of medium-sized local radio galaxies 
may be explained by gas masses of $10^9$ to $10^{10} \, M_\odot$ in massive elliptical
galaxies. Furthermore, the simulations provide a theoretical basis for the
observed correlation that emission line nebulae are generally found to be
brighter on the side of the shorter lobe in high redshift radio galaxies
\citep{McCarthy+1991}. This interaction of jets with the cold gas phase suggests
that star formation in evolving high redshift galaxies may be affected
considerably by jet activity.
\end{abstract}


\section{Introduction}

Asymmetries in double radio sources have been a matter of discussion since more
than 40 years, when \citet{RyleLongair1967} examined a conjectured relativistic
propagation. This discussion has included asymmetries both in brightness and in
length (e.g. of the lobes or the core--hotspot distance). The underlying
assumption mostly has been that jets are intrinsically symmetric objects as
they are launched near the central supermassive black holes. While relativistic
effects may cause brightness asymmetries by Doppler beaming
\citep{WardleAaron1997} and length asymmetries due to finite light travel times
\citep[examined by][]{Scheuer1995}, several studies
\citep{Pedelty+1989,McCarthy+1991,ArshakianLongair2000,Jeyakumar+2005} have
suggested that an inhomogeneous environment has a stronger impact on the
observed asymmetries. This is in agreement with \citet{Scheuer1995} finding only
a slow jet head propagation to be compatible with observations. In contrast to
the local universe, where low gas mass fractions are generally found in radio
galaxies, galaxies at high redshift show gas masses of the same order as the
stellar mass of the hosting galaxy \citep{Tacconi+2010}. Such large amounts of
gas in and around galaxies may have considerable effects on the jets. The
interaction of jets with the clumpy interstellar medium (ISM) and gas accreted
onto the galaxy may have a major impact on the formation of stars in the
galaxies and the galaxies' energy budget. These questions go much beyond the
merely morphological properties of the radio galaxies, but are clearly linked
with the kinematics, dynamics and evolution of the galaxy as a whole.
Observations of high-redshift radio galaxies commonly show the ``alignment
effect'' of having extended and sometimes very massive emission-line nebulae
aligned with the radio source \citep{McCarthy1993,Nesvadba+2008}. These nebulae
almost always are brighter in [OII] emission on the side of shorter jet. This
suggests that there may be vivid interaction going on: assuming the brighter
shock-excited gas traces the location of stronger interaction of the jet with
its environs, this may be an indication of the jet being responsible for the
formation of the emission line nebulae either directly or via its backflow
\citep*{Gaibler+2009,HLRS2010}.

Aiming towards understanding the interaction of jets with a complex multiphase
environment at high redshift, we have performed 3D hydrodynamic simulations of
a pair of jets with a surrounding clumpy gaseous disc. In this paper, we
present first results from the simulations, concentrating on jet asymmetries
caused by the clumpy ISM. We present results of Monte Carlo simulations based
on analytical approximations of the jet propagation through a clumpy ISM aside
with the full hydrodynamical simulations to examine the link between the
gaseous disc properties with the expected jet length asymmetries. A more
detailed examination of the simulations with respect to star formation and
feedback energetics as well as the kinematics of the dense gas will be
presented in a subsequent paper.

\section{The Model}

\begin{figure}
    \centering
    \includegraphics[width=0.9\linewidth]{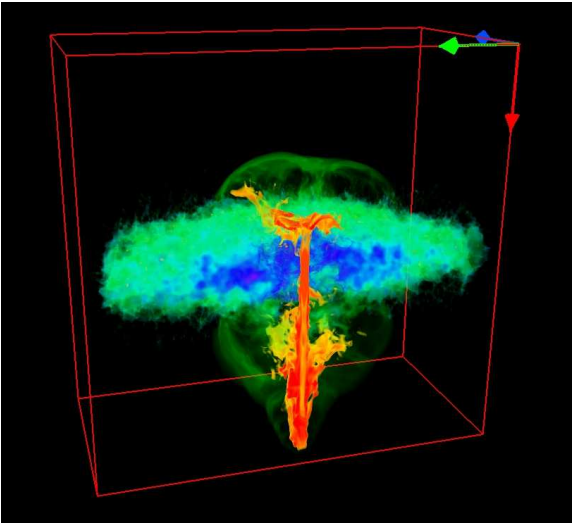}
    \caption{3D density volume rendering of the jet pushing through the dense
    gaseous disc at $t = 5.4$ Myr, box size $32 \times 32 \times 16$ kpc
    (cut in the plane of the jet). Low density is indicated by red,
    high density by blue.}
    \label{fig:jetdisk-3d}
\end{figure}
\begin{figure}
    \centering
    \includegraphics[width=0.9\linewidth]{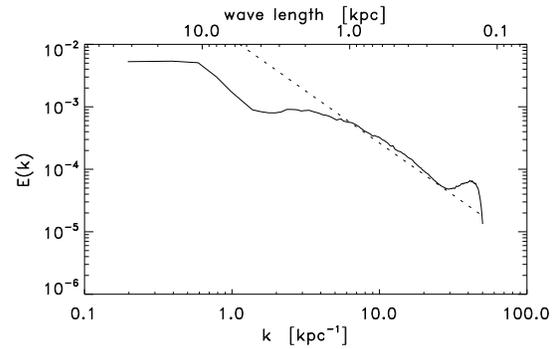}
    \caption{Energy spectrum $E(k)$ of the density field at the initial
    conditions.  The dotted line shows a power law $E(k) \propto k^{-5/3}$ for
    comparison. The bump at $k \sim 40$ kpc$^{-1}$ is a sampling effect since
    the disc cells are on a grid a factor of 2 coarser than the finest grid.}
    \label{fig:energyspectrum_initial}
\end{figure}

We set up a clumpy gaseous disc contained within a homogeneous hot atmosphere
(Fig.~\ref{fig:jetdisk-3d}). The atmosphere has density of $\rho_\text{a} =
0.05 \, \mp \, \mathrm{cm}^{-3}$ and a temperature of $T_\text{a} = 1.15 \times
10^7$ K. The log-normal density field of the disc is constructed from a
normal-distributed (mean $0$, standard deviation $1$) and clumpy field
$f(\vec{x})$, constructed in Fourier space, by
\begin{equation}
    \rho(\vec{x}) = \tilde{\rho} \, \exp \left\{ \sigma f(\vec{x}) \right\} \,
        \exp \left(-R/R_0 \right) \, \text{sech}^2 \left( h/h_0 \right) \, ,
\end{equation}
with the central median disc density $\tilde{\rho} = 10 \, \mp$ cm$^{-3}$ and
the disc scales $R_0 = 5$ kpc and $h_0 = 1.5$ kpc ($R, h$: cylindrical radius
and vertical height, respectively), using a temperature of $10^4$ K. The disc
values are only set where $\rho(\vec{x}) > \rho_\text{a}$, and a strict cutoff
is applied at $R = 16$ kpc and $h = 6$ kpc. The total mass of the gaseous disc
is $1.5 \times 10^{11} M_\odot$. For any small volume, the disc densities show
a log-normal distribution, but on the global disc scale, the imposed radial and
vertical profiles increase the contribution of lower densities. The standard
deviation $\sigma$ of the disc's log-density corresponds to turbulence of Mach
number 5 ($\sim 80$ km s$^{-1}$) \citep{Kritsuk+2007}, a value similar to large
evolving discs at high redshift \citep{FoersterSchreiber+2009}. The
two-point structure of the density field is described by the Fourier energy
spectrum $E(k)$ (Fig.~\ref{fig:energyspectrum_initial}), which follows $E(k)
\propto k^{-5/3}$ for large wave numbers $k$, but is intentionally damped
towards smaller wave numbers, avoiding large inhomogeneities on scales $l >
h_0$.
The initial configuration is evolved for 2 Myr, and after this initial
relaxation phase a bipolar (back-to-back) jet is introduced in the centre of
the disc. It is realized by a cylindrical orifice of jet plasma ($\rho_\text{j}
= 5 \times 10^{-5} \, \mp$ cm$^{-3}$, $v_\text{j} = 0.8$ c) with the same
pressure as the hot atmosphere and with a radius of $r_\text{j} = 0.4$ kpc and
an initial length in both directions of $3 \, r_\text{j}$, respectively. The
kinetic power of the jet is $L_\text{kin} = 5.5 \times 10^{45}$ erg s$^{-1}$.

This setup is evolved by the RAMSES 3.0 code \citep{Teyssier2002}, a
non-relativistic second-order Godunov-type adaptive mesh refinement (AMR) code.
The total computational domain extends over a cubic box of size 128 kpc, with a
coarse grid of $1$ kpc resolution and adaptive refinement down to $62.5$ pc
(effective resolution $2048^3$). We use the HLLC solver and the MonCen slope
limiter, refine on the entire disc and on 10 per cent gradients in density or
pressure or speeds above $0.1 \, c$, and assume $\gamma = 5/3$ for all phases.
This enables us to study virtually the entire jet--disc system at the finest
resolution.
We include the important effect of radiative cooling (as implemented in the
code) for a metallicity $Z = 0.5 Z_\sun$ \citep{Erb2008}, but exclude gravity
and stabilize the disc by imposing a minimum temperature of $10^4$ K since we
do not include stellar feedback, making the disc sufficiently stable over the
simulated time scale.

\section{Simulation Results}

\begin{figure*}
    \centering
    \includegraphics[width=0.99\linewidth]{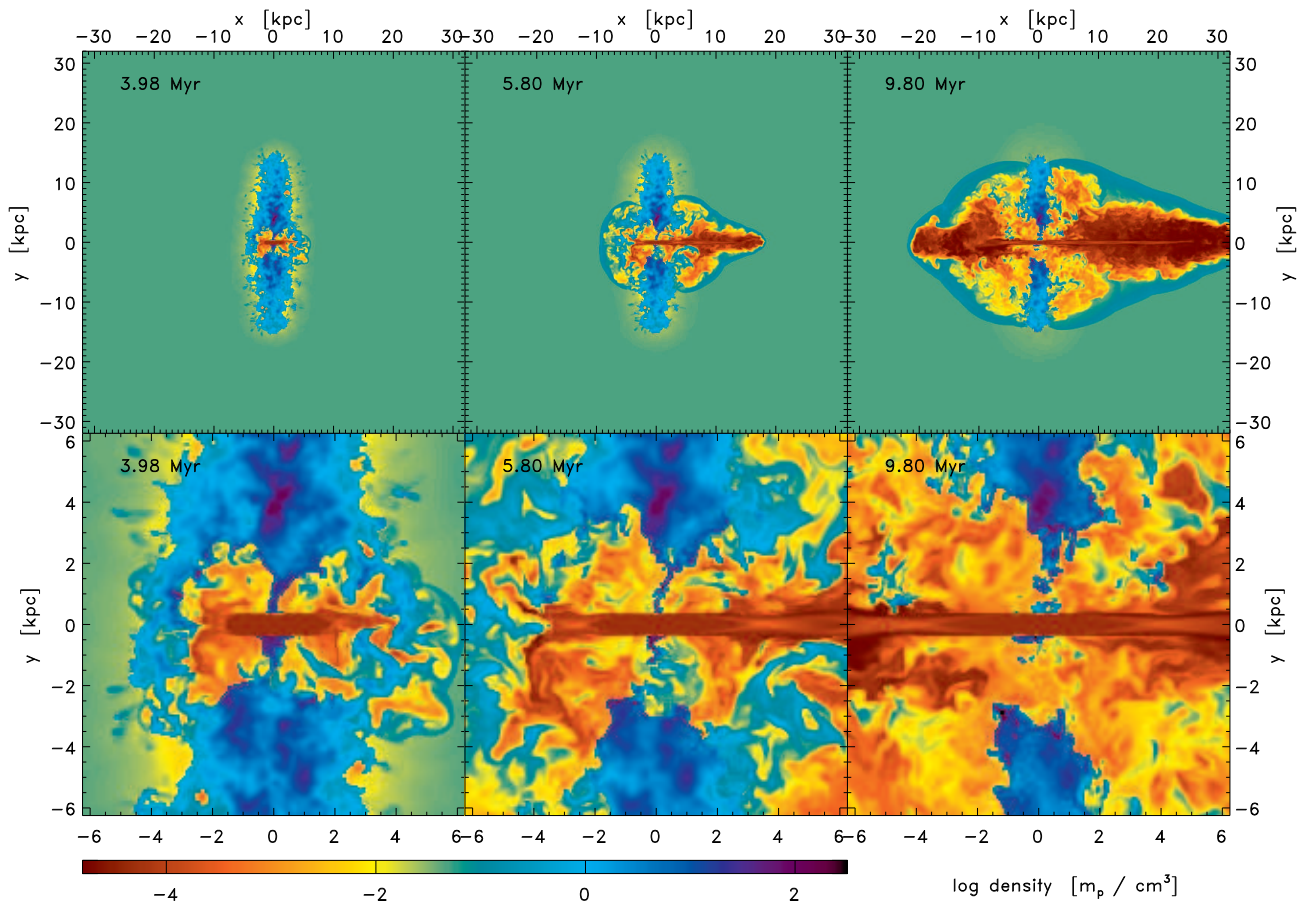}
    \caption{Density slices through the plane of the jet showing the evolution
    of the jet--disc interaction at three different times (\emph{left} to
    \emph{right}) and for the larger scales (\emph{top}) as well as the central
    region (\emph{bottom}).}
    \label{fig:evol-d-panel}
\end{figure*}
The jet inflates two strongly over-pressured cavities filled with diffuse jet
plasma in the central region of the disc and forms a disc-like structure of
strongly compressed gas in the midplane due to the synchronous start of the two
jets (the evolution is shown in Fig.~\ref{fig:evol-d-panel}). The gas in the
midplane of the disc at larger radii is largely unaffected by the jet due to
the high densities there, while the cavities expand in vertical direction and
form a pressure-driven blastwave in the outer layers of the disc and the
surrounding halo gas. The jet beam, however, is still restrained by dense
clumps and filaments. This early phase corresponds to the ``flood and channel''
and the ``energy-driven bubble'' phases described by
\citet{SutherlandBicknell2007}. At this stage, a remarkable asymmetry between
the two jets develops. The right-hand jet is only slightly deflected by a dense
clump in its way, which it pushes sideways and considerably ablates, and then
quickly pushes forward to the expanding bubble which it pierces at $t \approx
5$ Myr (``jet breakout phase''). The left-hand jet, however, is still
restrained by dense clumps at somewhat larger radii (middle plots in
Fig.~\ref{fig:evol-d-panel}), causing a delay in its propagation with respect
to the opposite jet. It reaches the expanding blastwave shell much later,
eventually piercing it at $t \approx 8$ Myr. Already in the earliest
(blastwave) phase, the jet's environment becomes asymmetric since the blastwave
is able to clear more of the environment on the right side than on the left
side, where a group of dense clumps makes this blastwave clearing less
effective. It is also worth noting that the restraining clouds are not simply
the ones located there already in the initial conditions, but they are
generally moved and reshaped by advection and compression caused by the
blastwave and the shocked beam plasma (``second order'' restraints). In case of
the left jet, one of these clouds was just in the middle of the jet's path.
During this phase, the jet plasma exhibits a complex velocity structure,
forming back-flowing vortices in some regions and forward-flowing streams and
plumes in others\footnote{Movies available online make this more evident. 
}. Once the jets have pierced the expanding shell, they show the classic
morphology known from jet simulations, propagating faster through the lower
density environment, although the lateral expansion of the cocoons is still
affected by the pierced shell.

To examine, whether this asymmetry could have been expected from the clumpy
disc set up in the initial conditions, we approximate the propagation of the
jet head by one-dimensional momentum balance for very underdense jets
\citep[e.g.][]{KrauseVLJ1} 
\begin{equation}
  v_\text{h} \approx \left( \rho_\text{j} / \rho \right)^{1/2} \: v_\text{j}
\end{equation}
and find a propagation time out to a distance $x$ of 
\begin{equation}
  t = \frac{1}{\rho_\text{j}^{1/2} v_\text{j}} \; 
    \int^x \rho(x^\prime)^{1/2} \: \text{d}x^\prime  \, .
  \label{eqn:timeestimate}
\end{equation}

\begin{figure}
    \centering
    \includegraphics[width=0.9\linewidth]{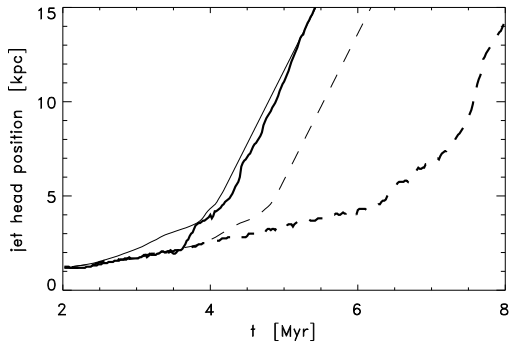}
    \caption{Momentum-balance estimate of jet propagation (\emph{thin lines},
    derived from initial conditions) and actual measurement (\emph{thick}) for
    the left (\emph{solid}) and right (\emph{dashed}) jet.}
    \label{fig:momentumbalanceevolution}
\end{figure}
Fig.~\ref{fig:momentumbalanceevolution} shows the analytical estimate along
with the measured position of the jet head based on the initial density field.
The right jet is described quite well by the estimate, while the left one needs
considerable additional time to break out compared to the expected behaviour
($2.7$ Myr vs. $0.7$ Myr expected from momentum balance). We attribute this to
the mentioned ``second-order'' restraints -- changes of the surrounding ISM
caused by the jet activity itself. They are small at early times, but become
important later once the jet has propagated through most of the disc and hence
changed its environment significantly from the initial configuration. The
asymmetry of the jets is also understandable in light of the initial disc mass
in the jet beam's path, which is about $2\times 10^7$ M$_\odot$ for the right
jet and $4\times 10^7$ M$_\odot$ for the left jet.

\section{Monte Carlo Simulations of Disc-induced Asymmetries}

We here investigate how the disc properties affect the distribution of
asymmetries in a statistical sense, by generating a large ensemble of discs.
Which disc leads to which typical asymmetries? Once the jets have broken out of
the disc or any other clumpy ISM, the major effect on the jet morphology will
be a propagation delay between the jets on both sides: in general, one jet will
have propagated already somewhat further than the other. Correspondingly, an
absolute length asymmetry is expected rather than a relative length asymmetry
as in the case of asymmetries caused by light travel time
\citep{ArshakianLongair2000}. Due to the computational demands, a statistical
treatment of this propagation delay by means of  3D hydrodynamic simulations is
hardly possible. However, despite the considerable differences between the 1D
estimate and the true propagation caused by individual clumps, simple momentum
balance works well enough to obtain first order estimates of the effects of a
clumpy interstellar medium on the jet propagation and the resulting arm length
asymmetries, as well as the impact of disc parameters on it. Clearly, one has
to keep in mind that the ``second-order'' effects may in general cause even
larger ranges of asymmetries than expected from a 1D estimate. 
We note that while we have assumed a
fully collimated, non-relativistic jet for our estimates, the Monte-Carlo
approach in combination with the one-dimensional momentum balance could also be
used for a conical and/or relativistic jet, which may be more realistic for
modelling the innermost regions, and any other ISM density structure.
\begin{figure}
    \centering
    \includegraphics[width=0.9\linewidth]{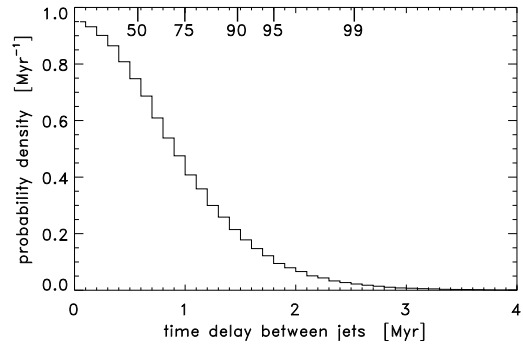}
    \caption{Probability density function derived from a Monte Carlo run based
    on the 1D momentum balance estimate for $\tilde{\rho} = 10$ m$_\mathrm{p}$
    cm$^{-3}$, $\sigma = 1$, $h_0 = 1.5$ kpc and the energy spectrum used for
    the hydrodynamic simulation. The locations of a set of percentiles are
    indicated by lines attached to the upper coordinate axis.}
    \label{fig:mcpdf}
\end{figure}

\begin{figure*}
    \centering
    \includegraphics[width=0.49\linewidth]{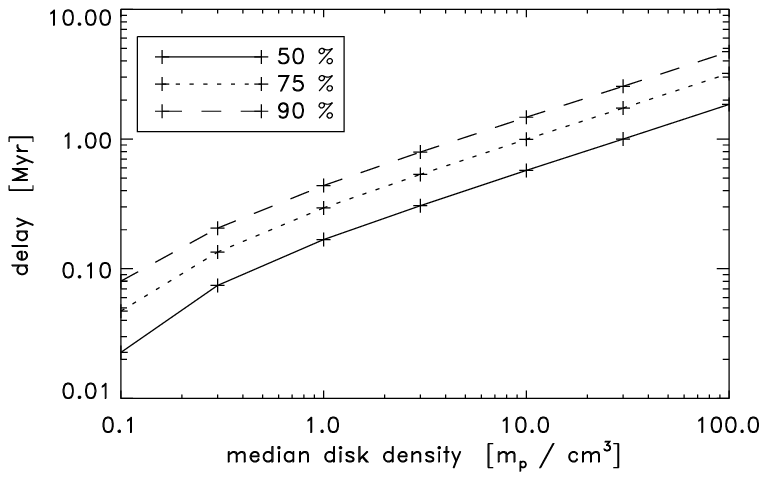}
    \includegraphics[width=0.49\linewidth]{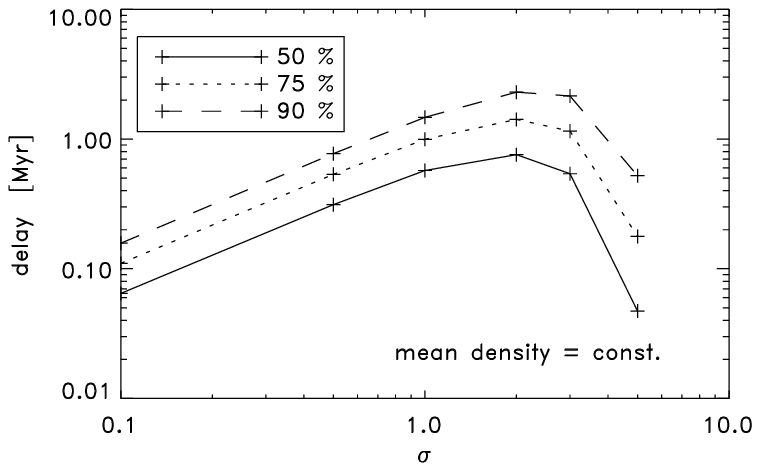}
    \includegraphics[width=0.49\linewidth]{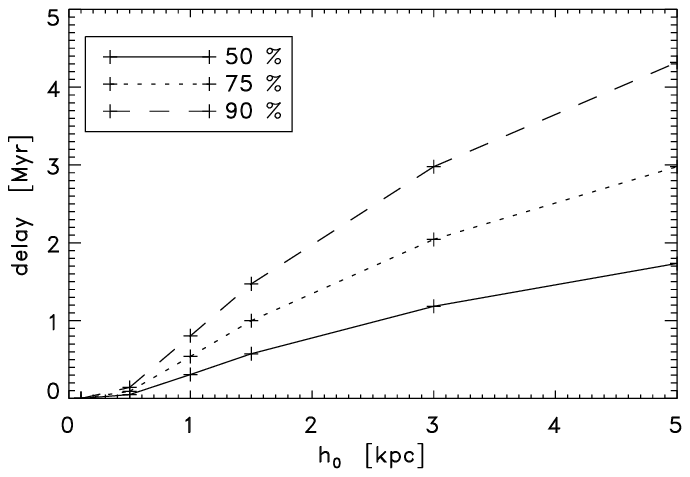}
    \includegraphics[width=0.49\linewidth]{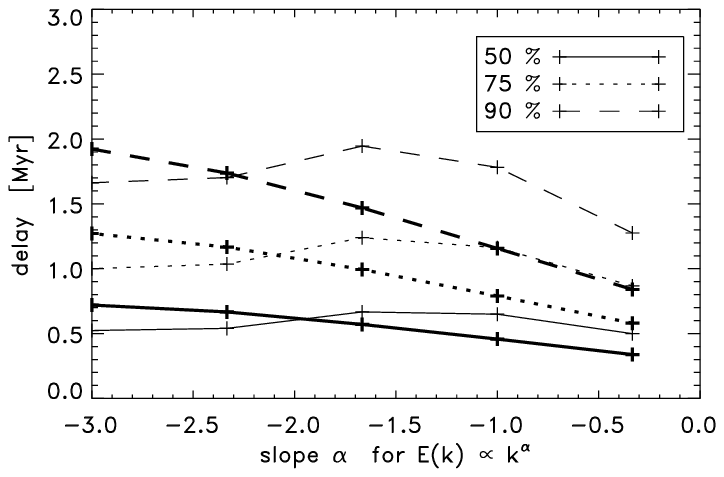}
    \caption{Impact of ISM disc properties on the distribution of the jet
    propagation asymmetry derived from the Monte Carlo simulations, plotted for
    the 50, 75 and 90 percentiles. \emph{Top left:} disc density
    $\tilde{\rho}$, \emph{top right:} density variations $\sigma$ (related to
    ISM turbulence), \emph{bottom left:} disc scale height $h_0$, \emph{bottom
    right:} Fourier energy spectrum shape and slope $\alpha$ for large wave
    numbers s of jets in terms of time delays for constant (for the latter,
    \emph{thin lines} are for pure power law energy spectra, \emph{thick lines}
    are for the case with damped small wave numbers as in the simulation
    initial conditions).
    \label{fig:mc-parameters}
    }
\end{figure*}

To examine the impact of disc parameters on the asymmetries caused by the disc,
we performed Monte Carlo simulations of the jet propagation using the momentum
balance estimate above. We varied the median value $\tilde{\rho}$ and the
log-normal distribution width $\sigma$ of the disc density, the disc scale
height $h_0$, the power spectrum shape and its slope at large wave numbers
(Figs.~\ref{fig:mcpdf} and \ref{fig:mc-parameters}). The asymmetries are
described by a time delay -- the time difference between both jets once the
disc has been left behind. A typical jet head propagation speed in the ambient
medium would allow to relate this to an asymmetry in length. Since a homogeneous
distribution is used for the diffuse ambient gas, no further asymmetries are
caused at a later time within this statistical model and the asymmetries are
entirely due to the clumpy disc. Fig.~\ref{fig:mcpdf} shows the probability
density for the fiducial values of the hydrodynamic simulation. The time
asymmetry of $0.7$ Myr seen in Fig.~\ref{fig:momentumbalanceevolution} for the
1D propagation model of our hydrodynamical setup lies well between the 50 and
75 percentiles (cumulative probability) and hence can be considered a
``one-$\sigma$ event''. In the following, we concentrate on the effects of the
mentioned parameters on the 50, 75 and 90 percentiles of the delay
distribution. Only one parameter is varied at a time while the others have
their fiducial value.

The mean density of the disc has a large impact on the jet asymmetries. For
median densities much larger than the diffuse background gas, the delays vary
$\propto \tilde{\rho}^{1/2}$, which is understandable from
\eqref{eqn:timeestimate}; clearly, this also translates directly into changes
of the total disc mass.
The density variation within the disc, described by the log-normal distribution
width parameter $\sigma$, can be related to the turbulent Mach number
$\mathcal{M}$ of the disc by
\begin{equation}
  \sigma^2 = \ln \left( 1 + b^2 \mathcal{M}^2 \right)
  \label{eq:Kritsuk-M-sigma}
\end{equation}
with $b \approx 0.26$ \citep{Kritsuk+2007}. For the effects of $\sigma$, it is
important to consider that for a log-normal distribution, the median value
$\tilde{\rho}$ relates to the mean density as $\langle{\rho}\rangle =
\tilde{\rho} \: \exp(\sigma^2/2)$, resulting in an increasing mean density and
hence total disc mass $M = 4 \pi R_0^2 h_0 \langle{\rho}\rangle$ for increasing
$\sigma$ and constant $\tilde{\rho}$. To avoid effects of changing density, the
top-right plot in Fig.~\ref{fig:mc-parameters} displays the changes in delay
for constant mean density (i.e. constant disc mass; the median density changes
along with $\sigma$). It shows an increase in delay until $\sigma \approx 2$
where it then decreases again. While the former is a result of a more and more
inhomogenous disc density distribution, a very wide range of densities (for a
highly turbulent medium with large $\sigma$) results in a small median density
and only small filling factors taken by gas of densities larger than the
background medium value and thus lower probabilities for clumps being in the
jet's path (for a constant mean density, the volume filling factor of dense
clumps in a log-normal density distribution decreases $\propto
\exp(-\sigma^2/8)/\sigma$ for increasing $\sigma$). For a constant median
density, however, the delay times would increase even stronger for $\sigma
\gtrsim 2$ due to the enormously increasing total disc masses. Interestingly,
the delays for the wide range of turbulent Mach numbers 2 to 30 (corresponding
to $0.5 < \sigma < 2$) are only weakly dependent on the exact value of
turbulence. The disc scale height $h_0$ shows a near-linear behaviour of the
delays with some deviations for values $h_0$ similar to the initial jet length,
which are an artefact of not modelling the earliest phase of jet propagation.
For zero initial jet length, the ``convergence point'' at $h_0 \approx 0.5$ kpc
is located at the origin. In contrast to the previous dependencies, the effect
of the shape and large wave number powerlaw slope $\alpha$ of the Fourier
energy spectrum $E(k)$ is quite small since even without damping of small wave
numbers these are suppressed by the imposed vertical profile scale $h_0$.
Furthermore the effects of small scale perturbations (stronger for flatter
energy spectra, $\alpha \to 0$) are increasingly averaged out over the length
scale of $h_0$. We conclude from the Monte Carlo simulations that the
distribution of disc-induced asymmetries mostly depends on the average disc
density or total mass and to a weaker extent on the density variations within
and scale height of the disc for plausible values of the latter two.

\section{Discussion \& Conclusion}
For a given propagation delay and once the jet has propagated beyond the
vertical disc/ISM boundaries, the observed length asymmetry will in general be
influenced by the density profile of the ambient medium, which itself governs
the propagation of the hotspots and lobes. Double radio sources in this stage
are expected to exhibit a length asymmetry $\Delta l = v_h \, \Delta t$
dependent on the propagation delay $\Delta t$ due to the clumpy medium and the
typical jet propagation speed $v_h$ outside. The absolute length asymmetries
due to the ISM should be roughly independent of the source size if the density
profile doesn't lead to a large change in $v_h$. We have compared this to the
observed asymmetries in radio galaxies, using the 3CRR sources with with FR II
morphology and $z < 1$ from \citet{Mullin+2008}\footnote{available at
http://zl1.extragalactic.info} for a $\Lambda$CDM cosmology with
$\Omega_\Lambda = 0.7$, $\Omega_\text{m} = 0.3$ and $h = 0.7$. Of these, only
sources with clearly identified hotspots on opposite sides are considered (88
sources) and projection effects are not accounted for here (differences should
be less than 50 per cent for most radio galaxies). Although the relative
asymmetries (with respect to the total source size) decrease for larger sources
\citep{ArshakianLongair2000}, the absolute arm length asymmetries increase
much.
\begin{figure}
    \centering
    \includegraphics[width=0.9\linewidth]{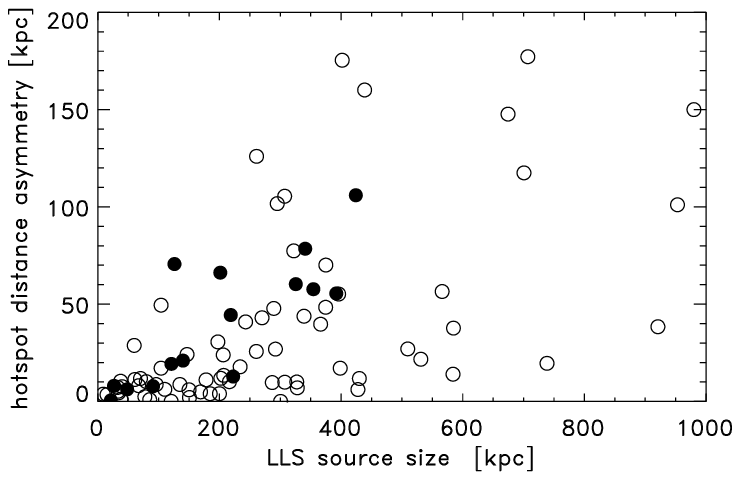}
    \includegraphics[width=0.9\linewidth]{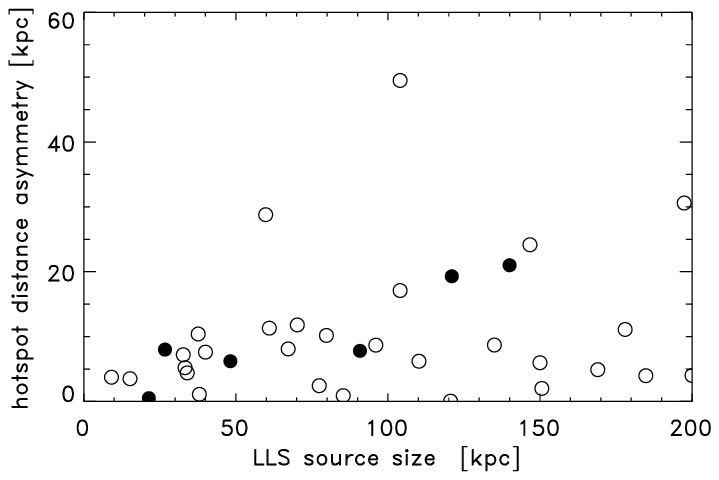}
    \caption{Arm length asymmetries for the chosen sample from
    \citet{Mullin+2008}: large (top) and only small (bottom) sources. Radio
    galaxies are represented by open circles, quasars by filled circles.}
\end{figure}
For the smaller sources (sum of LLS $\le 100$ kpc, 20 sources), however, the
scatter does not vary strongly with size and we argue that asymmetries caused
by a clumpy ISM cannot be larger than the scatter seen on these scales (mostly
between $0$ and $\sim 10$ kpc). The corresponding percentiles are $7.6$ kpc (50
per cent), $10.2$ kpc (75 per cent) and $11.8$ kpc (90 per cent). Based on a
fiducial jet propagation speed of $0.05 \; c$, the corresponding percentiles
for the propagation delay are $0.50$, $0.66$ and $0.77$ Myr. Comparison of
these values with our Monte Carlo runs now rises on the question, how typical
situations as in our hydrodynamical simulation with asymmetries larger than
expected are. 
If this is the rather typical case, and actual asymmetries may be factors of a
few larger than the 1D estimates, the observed typical asymmetries could be
described by asymmetries in the Monte Carlo runs of $0.1$--$0.2$ Myr, suggesting
disc densities are a factor of $>10$ smaller than in our chosen set of
parameters (with other combinations of parameters possible, or course). Although
a more detailed comparison with observed data is beyond the scope of this paper,
the corresponding masses for a dense clumpy medium of $10^9$ to $10^{10}
M_\odot$ may be compatible with actually measured gas masses \citep[as e.g.
in][]{Catinella+2010,Emonts+2010}. However, if the asymmetric clearing of
the central region by the initial blastwave and clumps remaining in the jet's
path are rather rare, the observed asymmetries require disc densities almost as
high as in our hydrodynamical setup even for the local universe, with total
masses of $10^{10} M_\odot$ and more. To us, the first case seems more probable
since the asymmetric blastwave simply is caused by the asymmetric initial
density and hence no exceptional behaviour.

While the asymmetries of small radio sources (below 100 kpc) may be explained
by a clumpy ISM, this is not true for the larger sources with larger absolute
length asymmetries. Although propagation of delayed jets through a declining
density profile can result in increasing absolute asymmetries, it is hardly
possible that they grow by a factor of 10 or more. These large sources might
actually become more asymmetric by a strongly asymmetric ambient density
profile at large scales \citep{Jeyakumar+2005} or due to an unstable jet
propagation on one side possibly caused by instabilies or obstructing clumps in
one jet. For the largest sources with large propagation speeds caused by a very
low density environment, light travel time effects may actually play a
significant role even if it doesn't for the smaller sources.

\begin{figure}
    \centering
    \includegraphics[width=0.7\linewidth]{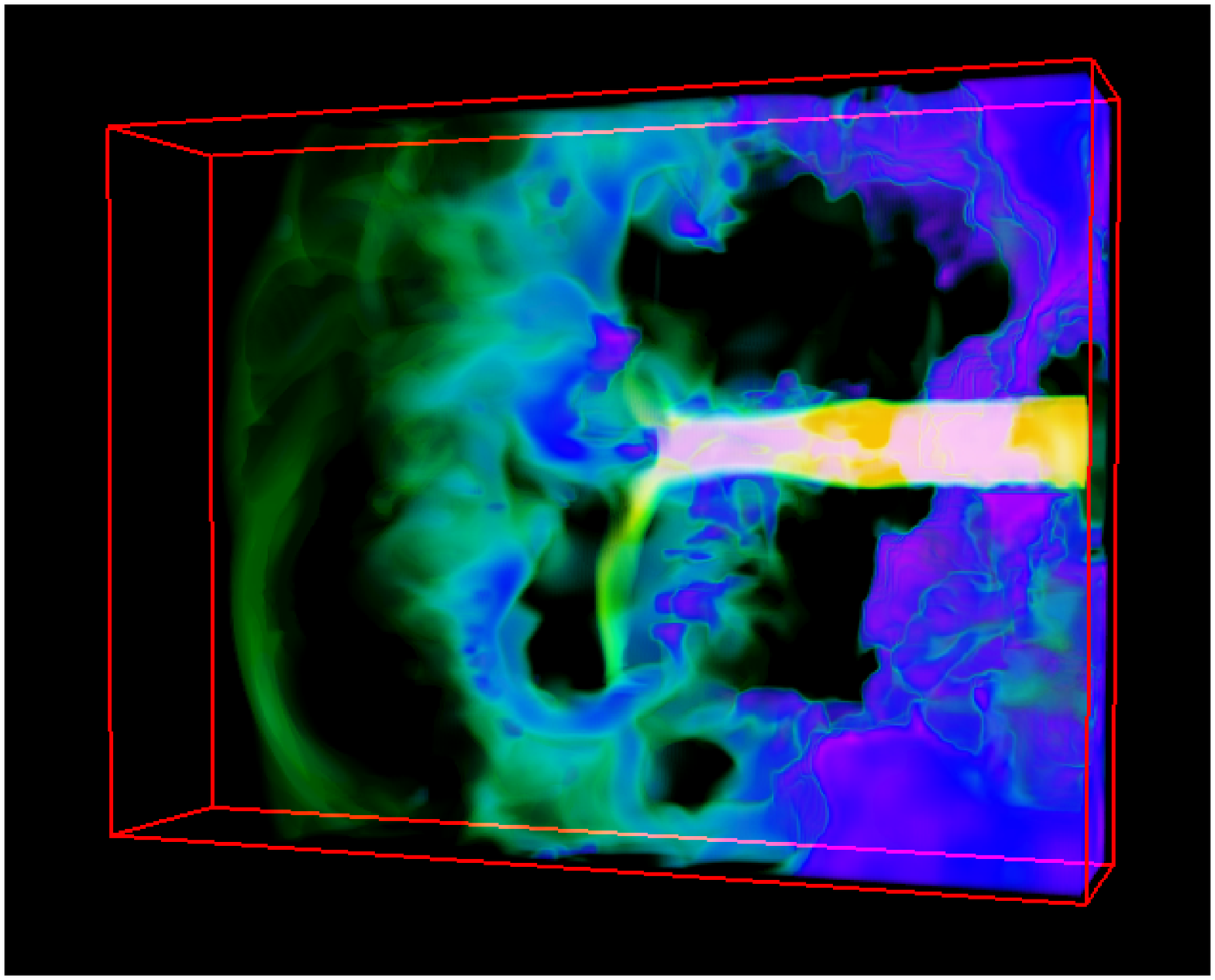}
    \includegraphics[width=0.7\linewidth]{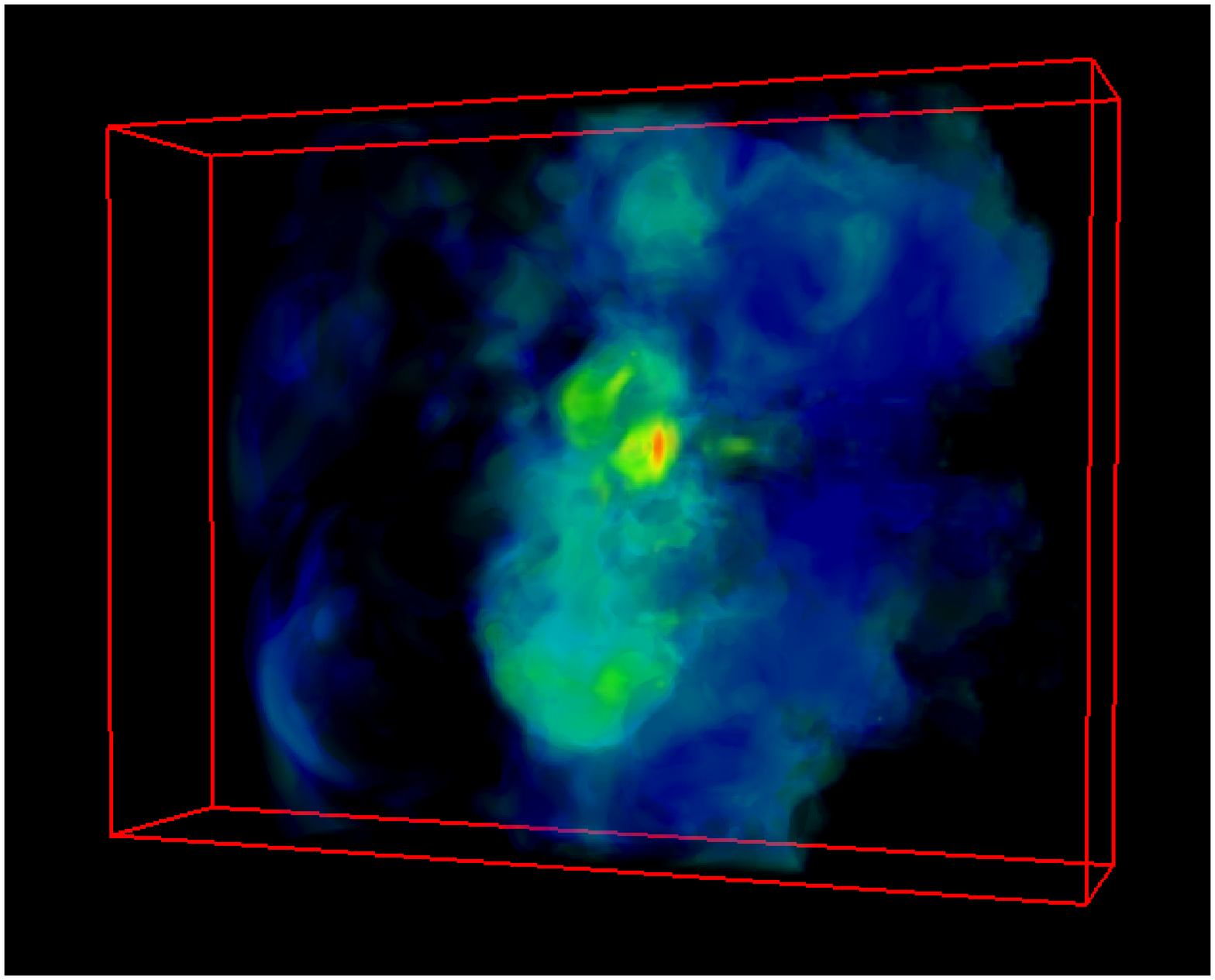}
    \caption{3D Volume rendering of the interaction of jet with a group of
    clumps that caused the additional delay in the hydrodynamic simulation
    compared to the 1D propagation estimate. \emph{Top:} density, with jet
    plasma tracer overlaid in yellow. High densities are blue, lower densities
    green. \emph{Bottom:} pressure, with high pressures in red and low pressures
    in blue. The boxes are $8.8$ kpc wide and $1.8$ kpc thick, $t = 4.84$ Myr.}
    \label{fig:denseclumps}
\end{figure}
\citet{McCarthy+1991} have found that in almost all cases of their sample, that
the extended optical line emission in [OII] is brightest on the side of the
shorter lobe, indicating that environmental effects cause the length
asymmetries in powerful radio sources. Our 3D hydrodynamical simulations are
now able to provide theoretical support for this scenario. Dense clouds or
atomic or molecular gas in the ISM are able to restrain a propagating jet for
some time due to their large inertia, causing asymmetries in the observed radio
source. At the same time, the impact of the jet has considerable effects on the
clouds themselves: depending on the exact location and cloud properties, they
are accelerated, shock-ionized and ablated by the jet's ram pressure and
compressed by the high pressure in the jet cocoon. Cooling by line emission
stabilizes them additionally \citep[cf.][]{Mellema+2002} and results in bright
emission at the location of the strongest interaction with the jet.
Fig.~\ref{fig:denseclumps} shows the density of the group of clumps responsible
for the large asymmetry in the simulation and the strongly increased pressure
(hence strongest expected shock ionization) at the impact location of the jet
on the clumps. The long-term survival of the clouds, however, does not only
depend on their direct interaction with the jet and its bow shock, but also on
their long exposure to the fast and turbulent cocoon plasma
\citep{KrauseAlexander2007}, which may destroy but also create emission line
clouds.

While radio sources in the local universe can grow more easily due to small gas
fractions and deposit their energy also at larger distances (e.g. stopping
cooling flows), radio galaxies at high redshift can be expected to show
considerably more interaction with the gas-rich and clumpy environment and the
cold gas phase. In fact, the clumpy appearance of high redshift galaxies
\citep{Elmegreen+2007} and clumps formed by Toomre unstable discs in
cosmological simulations \citep{Ceverino+2010} support this scenario. And while
strong radio source asymmetry is only the morphological consequence, the actual
feedback caused by the interaction may be vigorous.

\section{Summary}

We have performed 3D hydrodynamic simulations of powerful extragalactic jets
interacting with a disc-shaped clumpy interstellar medium of log-normal density
distribution. As a first step of analysis, in this paper we have focussed on
the jet asymmetry and its relation to the inhomogeneous ISM. The interaction
between the jet and its environment does not only occur by direct impact of the
jet beam on the clouds, but also by a early blastwave phase which considerably
changes the ISM and thereby changes the environment through which the beam
propagates then later-on. We find that the inner region is mostly cleared of
dense clumps by the jet activity except for some highly compressed structures
in the disc plane, while the gas at larger radii is only effected by the bow
shock and the jet backflow impacting on the disc vertically. Asymmetries are
caused by dense clouds that are in or near the jet's path and greatly
decelerate the jet head and result in a propagation delay which will generally
not be the same for both jets. We have analyzed the observed asymmetry by means
of Monte Carlo simulations based on a 1D propagation model for the jet through
the dense medium and find that the properties of the ISM can be translated to a
probability distribution of jet arm length asymmetries. Disc density and height
are found to have the largest effect on the asymmetry for realistic parameter
ranges. The shape and slope of the Fourier energy spectrum of the ISM density
only has a small effect on the asymmetry, as has the width of the density
distribution (related to the turbulent Mach number) due to the changes in the
filling factor. The hydrodynamic simulations showed that asymmetries in general
may have a wider distribution than expected from the 1D model due to the
complex interaction of the jet and its bow shock with gaseous clumps, going
much beyond simple energy disposal. According to our model, observed asymmetries 
of medium-sized local radio galaxies can be explained by gas masses in 
massive elliptical galaxies of order $10^9$ to $10^{10} \, M_\odot$. 
Furthermore, the hydrodynamic simulations provide a theoretical basis for the
observations of \citet{McCarthy+1991} that emission line nebulae in high
redshift radio galaxies are generally found to be brighter on the side of the
shorter lobe due to stronger dissipation and a resulting delay in jet
propagation. This interaction of jets particularly with the cold gas phase shows
that jet activity may have considerable impact on the evolution of high redshift
galaxies, where star formation may be triggered in compressed dense clumps of
gas, or quenched due to destruction or removal of the cold clumps.

\section*{Acknowledgments}
VG wishes to acknowledge financial support by the Deutsche
Forschungsgesellschaft (Priority Programme SPP 1177) and support by Romain
Teyssier with RAMSES. SK thanks Joe Silk for helpful discussions and acknowledges support from
the Royal Society Joint Projects Grant JP0869822. Computations were performed on the SFC cluster of TMoX at
MPE. The authors made use of VAPOR for visualization purposes
\citep{ClyneRast2005,Clyne+2007}.

\bibliographystyle{mn2e}
\bibliography{ref}

\begin{thebibliography}{}

\bibitem[\protect\citeauthoryear{Arshakian \& Longair}{Arshakian \&
  Longair}{2000}]{ArshakianLongair2000}
Arshakian T.~G.,  Longair M.~S.,  2000, \mnras, 311, 846

\bibitem[\protect\citeauthoryear{Catinella et~al.}{Catinella
  et~al.}{2010}]{Catinella+2010}
Catinella B. et~al.,  2010, \mnras, 403, 683

\bibitem[\protect\citeauthoryear{Ceverino, Dekel \& Bournaud}{Ceverino
  et~al.}{2010}]{Ceverino+2010}
Ceverino D.,  Dekel A.,    Bournaud F.,  2010, \mnras, 404, 2151

\bibitem[\protect\citeauthoryear{Clyne, Mininni, Norton \& Rast}{Clyne
  et~al.}{2007}]{Clyne+2007}
Clyne J.,  Mininni P.,  Norton A.,    Rast M.,  2007, New J. Phys, 9, 301

\bibitem[\protect\citeauthoryear{Clyne \& Rast}{Clyne \&
  Rast}{2005}]{ClyneRast2005}
Clyne J.,  Rast M.,  2005, in Erbacher R.~F., Roberts J.~C., Gr\"ohn M.~T., B\"orner K., eds., Visualization and Data Analysis 2005. Proc. SPIE, 5669, 284

\bibitem[\protect\citeauthoryear{Elmegreen, Elmegreen, Ravindranath \&
  Coe}{Elmegreen et~al.}{2007}]{Elmegreen+2007}
Elmegreen D.~M.,  Elmegreen B.~G.,  Ravindranath S.,    Coe D.~A.,  2007, \apj,
  658, 763

\bibitem[\protect\citeauthoryear{Emonts et~al.}{Emonts
  et~al.}{2010}]{Emonts+2010}
  Emonts B. H.~C. et~al., 2010, \mnras, 406, 987

\bibitem[\protect\citeauthoryear{Erb}{Erb}{2008}]{Erb2008}
Erb D.~K.,  2008, \apj, 674, 151

\bibitem[\protect\citeauthoryear{F{\"{o}}rster~Schreiber et~al.}{F{\"{o}}rster~Schreiber et~al.}{2009}]{FoersterSchreiber+2009}
F{\"{o}}rster~Schreiber N.~M. et~al.,  2009, \apj, 706, 1364

\bibitem[\protect\citeauthoryear{Gaibler \& Camenzind}{Gaibler \&
  Camenzind}{2010}]{HLRS2010}
Gaibler V.,  Camenzind M.,  2010, in Nagel W.~E.,  Kr{\"{o}}ner D.~B.,   Resch
  M.~M.,  eds, High Performance Computing in Science and Engineering {'}09.
Springer, p. 3

\bibitem[\protect\citeauthoryear{Gaibler, Krause \& Camenzind}{Gaibler
  et~al.}{2009}]{Gaibler+2009}
Gaibler V.,  Krause M., Camenzind M.,  2009, \mnras, 400, 1785

\bibitem[\protect\citeauthoryear{Jeyakumar, Wiita, Saikia \& Hooda}{Jeyakumar
  et~al.}{2005}]{Jeyakumar+2005}
Jeyakumar S.,  Wiita P.~J.,  Saikia D.~J.,    Hooda J.~S.,  2005, \aap, 432,
  823

\bibitem[\protect\citeauthoryear{Krause}{Krause}{2003}]{KrauseVLJ1}
Krause M.,  2003, \aap, 398, 113

\bibitem[\protect\citeauthoryear{Krause \& Alexander}{Krause \&
  Alexander}{2007}]{KrauseAlexander2007}
Krause M.,  Alexander P.,  2007, \mnras, 376, 465

\bibitem[\protect\citeauthoryear{Kritsuk, Norman, Padoan \& Wagner}{Kritsuk
  et~al.}{2007}]{Kritsuk+2007}
Kritsuk A.~G.,  Norman M.~L.,  Padoan P.,    Wagner R.,  2007, \apj, 665, 416

\bibitem[\protect\citeauthoryear{McCarthy}{McCarthy}{1993}]{McCarthy1993}
McCarthy P.~J.,  1993, \araa, 31, 639

\bibitem[\protect\citeauthoryear{McCarthy, van Breugel \& Kapahi}{McCarthy
  et~al.}{1991}]{McCarthy+1991}
McCarthy P.~J.,  van Breugel W.,    Kapahi V.~K.,  1991, \apj, 371, 478

\bibitem[\protect\citeauthoryear{Mellema, Kurk \& R{\"{o}}ttgering}{Mellema
  et~al.}{2002}]{Mellema+2002}
Mellema G.,  Kurk J.~D.,    R{\"{o}}ttgering H. J.~A.,  2002, \aap, 395,

\bibitem[\protect\citeauthoryear{Mullin, Riley \& Hardcastle}{Mullin
  et~al.}{2008}]{Mullin+2008}
Mullin L.~M.,  Riley J.~M.,    Hardcastle M.~J.,  2008, \mnras, 390, 595

\bibitem[\protect\citeauthoryear{Nesvadba, Lehnert, De~Breuck, Gilbert \& van
  Breugel}{Nesvadba et~al.}{2008}]{Nesvadba+2008}
Nesvadba N. P.~H.,  Lehnert M.~D.,  De~Breuck C.,  Gilbert A.~M.,    van
  Breugel W.,  2008, \aap, 491, 407

\bibitem[\protect\citeauthoryear{Pedelty, Rudnick, McCarthy \& Spinrad}{Pedelty
  et~al.}{1989}]{Pedelty+1989}
Pedelty J.~A.,  Rudnick L.,  McCarthy P.~J.,    Spinrad H.,  1989, \aj, 97, 647

\bibitem[\protect\citeauthoryear{Ryle \& Longair}{Ryle \&
  Longair}{1967}]{RyleLongair1967}
Ryle M.,  Longair M.~S.,  1967, \mnras, 136, 123

\bibitem[\protect\citeauthoryear{Scheuer}{Scheuer}{1995}]{Scheuer1995}
Scheuer P. A.~G.,  1995, \mnras, 277, 331

\bibitem[\protect\citeauthoryear{Sutherland \& Bicknell}{Sutherland \&
  Bicknell}{2007}]{SutherlandBicknell2007}
Sutherland R.~S.,  Bicknell G.~V.,  2007, \apjs, 173, 37

\bibitem[\protect\citeauthoryear{Tacconi et~al.}{Tacconi et~al.}{2010}]{Tacconi+2010}
Tacconi L.~J. et~al.,  2010, \nat, 463, 781

\bibitem[\protect\citeauthoryear{Teyssier}{Teyssier}{2002}]{Teyssier2002}
Teyssier R.,  2002, \aap, 385, 337

\bibitem[\protect\citeauthoryear{Wardle \& Aaron}{Wardle \&
  Aaron}{1997}]{WardleAaron1997}
Wardle J. F.~C.,  Aaron S.~E.,  1997, \mnras, 286, 425

\end{thebibliography}

\label{lastpage}

\end{document}